\documentstyle[12pt]{article}
\topmargin
-2.5cm
\textwidth
17cm
\textheight
25cm
\oddsidemargin
-0.25cm
\begin{document}
\baselineskip=3.0ex
\date{}
\title{ \large\bf {Standard Grand Unification from superstrings}}
\author{ {\large\rm {G. Aldazabal}} \thanks{Permanent Institutions: CNEA,
Centro
At\'omico Bariloche, 8400 S.C. de Bariloche, and CONICET, Argentina.}
\\ \\
\large\it {Departamento de F\'{\i}sica Te\'orica,} \\
\large\it {Universidad Aut\'onoma de Madrid,} \\
\large\it {Cantoblanco, 28049 Madrid, Spain.}
\\   \\}
\maketitle
\vspace{3in}

\begin{abstract}
Recent developments \cite{afiu,afiu2} about the construction
of standard $SO(10)$ and $SU(5)$ grand unified theories from
4-dimensional superstrings are presented. Explicit techniques
involving higher level affine Lie
algebras, for obtaining such stringGUTs
from symmetric orbifolds are discussed. Special
emphasis is put on the different constraints and
selection rules for model building in this string framework, trying
to disentangle those which are generic from those depending on the
orbifold construction proposed. Some phenomenological implications from
such constraints are briefly discussed.
\end{abstract}
\maketitle

\bigskip

\centerline{\it Talk at Susy 95, Paris, May 1995}
\newpage
\section{Introduction}

The motivations for considering the construction of GUTs in the framework of
string theory are manifold and rely on different basis or beliefs.
The main reason is undoubtedly the striking fitting of $sin^2\theta _W$ vs.
$\alpha _s $ when both supersymmetry and unification are assumed \cite{amal}.
Many of the virtues  emphasised by the first GUTs builders more than a decade
ago, are still appealing today. The fitting of SM content into simple GUT
multiplets (predicting that right handed quarks are SU(2) singlets etc.)
is an example.
Therefore it seems worth exploring the possibility of having  GUTs like models
in the context of a frame theory  for unifying gauge
and gravitational interactions. We will call this type of models StringGUTs
\cite{afiu}.

Let us mention, however, that some  other features signaled in GUTs
days appear less compelling in the presence of string theory.
Charge quantization for example may
be understood in this context from anomaly cancellation. Indeed, even the
attractive
idea of coupling constants unification could perhaps be achieved at the string
scale directly from Standard Model. Introduction of non fully
controllable features  would be required in this case (see Ref.\cite{iki}
for a discussion of this possibility).
Nevertheless, the main obstacle that remains is the lack of a  doublet-triplet
splitting mechanism (unless antrophic principle is called for fine tuning) for
avoiding fast proton decay. If StringGUTs are meaningful, they should furnish
at least a possible hint for solving this problem.

In building up StringGUTs we will not only be interested  in having
GUT groups like $SU(5)$ or $SO(10)$, but also
in their matter content (Higgses, fermion generations) and in the allowed
couplings indicating  possible symmetry breaking patterns, mass
generating terms, doublet triplet splitting terms etc., present at GUT scale.

The gauge group $G$ information in string theory relies on the presence of an
affine
Lie algebra at level $k$, realized by currents $J_a(z)$ ($a=1,\dots, rank G$)
living on the world sheet of
the string. Namely they satisfy
\begin{equation}
J_a(z)J_b(w)= \frac {k \delta _{ab}}{ (z-w)^2}+
\frac {if_{abc}J^c}{ (z-w)}+\dots
\label{ala}
\end{equation}
where $f_{abc}$ are the structure constants of the group Lie algebra.
Whereas in ten dimensions the only consistent groups are $E_8 \times E_8$ or
$SO(32)$ realized at level $k=1$, in four dimensions the situation is much
less constrained. The gauge group generically appears as a product of
non-abelian gauge groups $G_i$, realized at levels $k_i$,
 $ G_1 \times G_2 \times \cdots$  times $U(1)$ factors. However,
the values of levels are not completely arbitrary. In fact, vanishing of
conformal
anomaly requires  $k \le 7$ for $SO(10)$ , $k \le 4$ for $E_6$
or the more relaxed condition $k\le 55$ for $SU(5)$.
These are upper bounds; actual computations in specific models prove to be much
more restricted.
The methods proposed below, for example, cannot produce levels beyond
$k=3$ (and probably not bigger than $k=2$) for the GUT groups we are
interested in.
In particular, most of the models considered in the literature are normally
obtained at level $1$.

The values attained by level $k$ are very important, since they impose
(unitarity) limits on possible representations allowed. In
particular for level $1$, only vector or spinor representations are admitted
in $SO(2N)$  and only the fundamental or $10$ (and their conjugates) for
$SU(5)$ group.

In supersymmetric GUTs, it is crucial to have quiral fields
in the adjoint (or bigger representations) in order to achieve the breaking
of the GUT symmetry down to the Standard Model.
In the context of $N=1$ four
dimensional strings, this requires GUT groups realized at levels $k \ge 2$.
Therefore, building higher level gauge
groups becomes a necessary task. This problem was addressed in orbifold models
in Ref.\cite {fiq}.
In Ref.\cite {lew} it was considered for fermionic string
constructions and recent GUT model building attempts in this context
may be found in \cite{otros}.

In our specific work we will be mainly concerned with
symmetric orbifolds
(see \cite{afiu,afiu2} for a more detailed list of references.)
 This type of construction
is not better, in principle, than other four dimensional strings
pictures. The main advantage is, perhaps, that models are easy to handle,
in the sense that many desired aspects can be controlled without
need of computers. Furthermore, many consistency checks are available and
 world-sheet supersymmetry (which  may be a problem in fermionic
constructions \cite{cchl}) is ensured from the beginning.

 Before getting into these orbifold models, let us first briefly
discuss another stringent
stringy constraint for low energy model building. When considering
effective theories, only massless string states will be relevant. The nature
of this constraint relies on the conformal structure of the theory and may
be stated in simple words by saying that quantum numbers
do weigh (conformally). Thus, the more quantum numbers a particle
has, the less likely it is for it to be in the massless spectrum.
This is summarized in the following mass formula
\begin{equation}
{1\over 8}M_L^2\ =\
h_{KM}\ +\ h_{int}\ -1  \ .
\label{ml}
\end{equation}
where $h_{int}$ is the contribution to the mass of the particle from
the internal (compactified) sector and
$h_{KM}$ is the contribution of the gauge sector
to the conformal weight of the particle. For a state in a representation $R$
of a given non abelian gauge group it is given by
\begin{equation}
h_{KM}\ =\ \sum _i {{C(R_i)}\over {k_i+\rho_i}}
\label{peso}
\end{equation}
where $C(R)$ is the quadratic Casimir of the representation $R$, thus growing
with the size of $R$, illustrating what we said above. The allowed massless
representations must then satisfy $h_{KM} \le 1$. For GUT groups
$SO(10)$ and $SU(5)$, at levels $k=1,2$ they are given in Table 1.
\begin{table}
\begin{center}
\begin{tabular}{|c|c|c|c|c|c|c|}
\hline
${\bf SU(5)}$
   & ${\bf 5}$    &  ${\bf 10} $  &  ${\bf 24 }$ &  ${\bf 15}  $ &
 ${\bf 40}$  & ${\bf 50}$  \\
\hline
${\bf k=1}$ &  2/5  &  3/5   &  -  &  -  &  -  &  - \\
\hline
${\bf k=2  }$
& 12/35        &  18/35   &   5/7  &   28/35   & 33/35  & - \\
\hline
${\bf SO(10)}$
& ${\bf 10}$   &  ${\bf 16} $  &  ${\bf 45 }$ &  ${\bf 54}  $ &
 ${\bf 120}$ & ${\bf 126}$ \\
\hline
${\bf k=1}$ &  1/2  &  5/8   &  -  &  -  &  -  &  - \\
\hline
${\bf k=2}$ &  9/20 &  9/16  &  4/5   &  1      & -   & - \\
\hline
\end{tabular}
\end{center}
\caption{ Conformal weights $h_{KM}$ for different representations
of the unifying groups $SU(5)$ and $SO(10)$.}
\label{tdosa}
\end{table}
Therefore, some relevant phenomenological information may be already extracted
from very general considerations.
For example, a missing partner mechanism (which needs  $50 -plet $)  is not
allowed for in $SU(5)$, a
 $\underline {126}$ in
$SO(10)$ cannot be used for producing fermion mass matrices or see-saw
mechanism etc..
It is also worth noticing that the $\underline {54}$ representation in
$SO(10)$ is rather
peculiar. Its conformal weight being $h_{54}=1$ tells us  that
it cannot be charged with respect to other gauge groups and that
the internal right handed part of its vertex operator is trivial.
Couplings with  other states will therefore be quite restricted.
Moreover, a $\underline {54}$ can
only live in a sector with vanishing internal energy. For symmetric orbifolds
this sector must be one (an order two) of the untwisted sectors and
consequently, from orbifold
selection rules, it can be shown that self couplings are not
allowed for. These are  indications (which can be made more explicit in
specific
constructions) that a $\underline {54}$ behaves like a string modulus
\cite{afiu}.

The starting point in our explicit StringGUTs construction is (0,2) symmetric
orbifold of a four dimensional heterotic string. It proves convenient in this
case to deal with $Spin(32)$   instead of $E_8 \times E_8$ gauge lattice.
A replicated gauge group structure $G_{GUT}\otimes G'$ with $G_{GUT} \in G'$
 is then looked for by an adequate embedding of the orbifold action (as a shift
vector, a lattice automorphism etc.) into the gauge degrees of freedom.
The corresponding generators and levels are ($J$,$k=1$) and ($J'$, $k'=1$).
 The second step in our construction relies on the introduction
of a projection selecting only diagonal $J_a+J'_a$ combinations.
Thus, from eq. (\ref{ala}) and by noticing
that generators of the replicated groups commute among themselves
($JJ' = 0$),
we see that the diagonal group $G_{GUT}^D$ emerges at level $k+k'=2$ as
desired.
Of course, the introduction of such a projector implies the simultaneous
 appearance of twisted sectors (and related constraints), in order to maintain
modular invariance (see Ref.\cite {afiu2} for a detailed discussion of
constraints).

Notice that when the level 2 diagonal group is selected, a coset structure
emerges,  contributing with the missing conformal structure.
Consider the interesting
case of $SO(10)\times SO(10)$ with central charge $c=10$
broken to the diagonal group $SO(10)_2$ at level 2, with central charge
$c_2=9$.
The missing unity of central charge is provided by  the coset
$\frac {SO(10) \times SO(10)}{SO(10)_2}$. This is a rather peculiar coset. Its
 charge being one, it must be equivalent  to a free (orbifoldize)
compactified $S_1$ boson
\cite{ginsparg}. In fact it corresponds to $S_1/Z_2$ at the compactifying
radius
$\sqrt{\frac {5}{2}}$.

Let us stress that the general basics of  the methods discussed
above, rather than being  exclusive of orbifolds, are quite general.
They could be implemented, for example, in $N=2$ coset
constructions like Gepner \cite{gepner} or Kazama-Suzuki \cite{kz}
models, by embedding an order two internal modding as a permutation of
the replicated GUT groups. The replicated structure could be obtained by
embedding a shift into the gauge lattice as explained in Ref. \cite{fimqr}.
In this case, $SO(26)$ rather than $SO(10) \times E_8$ might be probably
 preferred.

The breaking to $G_{GUT}^D$ in orbifold models may be achieved through the
methods proposed in Ref. \cite{fiq}. Method (I) exploits the fact that, when an
orbifold
twist is embedded into the gauge degrees as an automorphism of the gauge
lattice- in this present case used to obtain the
structure $G_{GUT}\otimes G'$- non quantized Wilson line are allowed for. They
are then subsequently  added in order to continuously break to  level 2,
diagonal
GUT group. In the second method (II) the replicated structure is usually
achieved
through a shift and last breaking is realized by modding by a $Z_2$ which
permutes $G_{GUT}$ and $G_{GUT}'$ inside $ G'$.
The third method is field theoretical. The diagonal group is
obtained by looking for flat directions of the effective superpotential.
Let us mention that even if these procedures look quite different, in many
cases the same model at $k=2$ is obtained.
Nevertheless, method (II) is in some aspects
more versatile.  For instance the permutation twist may be accompanied by
a discrete shift which may be used to pick up a 45-plet (it never appears in
method
(I) or (II)) of $SO(10)$ instead of a $\underline {54}$.

More explicit information may be obtained in the symmetric orbifold context
from  eq.(\ref{ml}).
In fact, now the internal energy can be explicitly computed in terms of the
vacuum energy $E_0= \sum _{i=1}^3\ {1\over 2}|v_i|(1-|v_i|)$ where $v_i$
$(i=1,2,3)$ are known twist eigenvalues for each twisted sector.
We learn that all level 2 representations in Table 1 are allowed
in the untwisted sectors. In twisted sectors a $45$ could only fit
in sectors with $v=1/4(0,1,1)$ or $v=1/6(0,1,1)$
(further analysis discards even this possibility), 24-plets of $SU(5)$ are
forbidden in sectors $Z_3$, $Z_4$, $Z_6'$ and $Z_8$ orbifolds.

 As an illustration of what we have been discussing above,
 consider the following  $SO(10)_2 \times SO(8) \times U(1)^2$ StringGUT model
 in a $Z_2 \times Z_2$ symmetric
orbifold. This model can be
obtained from the three methods discussed above. Consider method (II). A
first breaking to  $SO(10)\times SO(18) \times U(1)^2$ at level one is
achieved by embedding the orbifold twists $\theta$ of the first $Z_2$ as a
shift {\nobreak $A=1/2(11111 00000 0000 10)$}
into the gauge degrees of freedom, and
similarly for the other twist $\omega$, as a shift
$B=\frac 1{2} (00000 00000 0000 11)$.
The breaking to the diagonal group is then achieved by assigning to a second
order direction of the compactifying cubic lattice, a Wilson line $\Pi$ acting
as a permutation
of the first  two blocks of five gauge bosons coordinates,
\begin{eqnarray}
& \Pi(X_1,X_2,X_3,X_4,X_5,X_6,X_7,X_8,X_9,X_{10}|\dots,X_{16}) = \\
& (X_6,X_7,X_8,X_9,X_10;X_1,X_2,X_3,X_4,X_5|\dots,X_{16})
\label{}
\end{eqnarray}
This model has four generations. Its massless spectrum is presented in Table 2.
$Q$ and $Q_A$ are the corresponding $U(1)$ charges. $Q_A$ is anomalous, the
anomaly cancelling, as usual, through the Green-Schwarz mechanism.
%\newpage
\begin{table}
\begin{center}
\begin{tabular}{|c|c|c|c|}
\hline
$Sector $
& $SO(10)\times SO(8)$ & $Q$ &  $Q_A$   \\
\hline
$  U_1  $  &     (1,8) &   1/2   &   1/2  \\
\hline
& (1,8)   &    -1/2   &  -1/2  \\
\hline
$  U_2  $ & (1,8)   & -1/2 & 1/2  \\
\hline
& (1,8)    &   1/2  &  -1/2  \\
\hline
$   U_3  $ & (54,1)   & 0   &  0  \\
\hline
& (1,1)    &    0  & 0   \\
\hline
& (1,1)    &  0  & 1    \\
\hline
& (1,1)    &   1  & 0   \\
\hline
& (1,1)    &  -1  & 0   \\
\hline
& (1,1)    &  0  & -1   \\
\hline
$\theta$   & $3(16,1)$   &  1/4  &  1/4  \\
\hline
& $(\overline{16},1)$   &  -1/4  &  -1/4  \\
\hline
$\omega$   & $3(16,1)$   &  -1/4  &  1/4  \\
\hline
& $(\overline{16},1)$   &  1/4  &  -1/4  \\
\hline
$\theta\omega$ &    $ 4(10,1)$   & 0   & 1/2  \\
\hline
& $4(10,1)$   & 0   & -1/2  \\
\hline
& $3(1,8)$   & 0   & 1/2  \\
\hline
& $(1,8)$   & 0   & -1/2  \\
\hline
& $8(1,1)$   & 1/2   & 0  \\
\hline
& $8(1,1)$   & -1/2   & 0  \\
\hline
\end{tabular}
\end{center}
\caption{Particle content and charges.}
\label{ttres}
\end{table}
%& $SO(10)\times SO(8)$ & $Q$ &  $Q_A$   \\

Let us conclude with a brief summary of the constraints
and selection rules for GUT model building from superstrings. As we
mentioned, some of these are very general and rely
on the 4-d heterotic string framework. All terms in the effective
superpotential must have
$dim \ge 4$  (mass terms ), only  level 2 representations
presented in Table 1 are admitted, $\underline {54}$ ${\underline {54}}$$ X$ or
\break
${\underline {54}}$${\underline{54}'}X$  couplings, where $X$ is a singlet  are
forbidden.
Other rules depend  on the symmetric orbifold
construction and on the methods proposed for  level $2$ models construction.
It can be seen for instance that: 1) There is place for only one $\underline
{54}$ or alternatively one $\underline {45}$ $SO(10)$. They must lie in
an order two untwisted sector.
2)  As a consequence, no self couplings $\underline {54} ^n$ (or $\underline
{45}$)
 are possible. 3) Couplings ${\underline {45}}$ ${ \underline {45}} X$ are also
forbidden etc. An extensive
discussion of these rules will appear in Ref. \cite{afiu2}.

With this type of constraints, it seems quite difficult to achieve GUT
symmetry breaking while keeping the MSSM particle content needed for
direct coupling unification.
 Then we may say that in
general, extra  particles besides those in the MSSM spectrum will remain
massless. For example if $SU(5)$ breaking is achieved through the adjoint
 $\underline {24}$, the partners of  Goldstone bosons, which usually acquire
mass
through self couplings in the potential, will remain massless.   They
transform as $(8,1,0)+ (1,3,0)+(1,1,0)$ under $SU(3)\times SU(2)\times U(1)$
and there is no direct coupling  unification close to $10^{16}Gev$ with
this extra matter. It remains to be seen if this is a generic, unavoidable
feature, calling for the introduction of an intermediate scale, or if models
with the MSSM minimal particle content may be found.
Models such as those proposed
in \cite{hall}, (without self interactions, with no bigger than
$\underline {54}$
dimensional representations..)
  look quite close to string GUTs building requirements. However some (from
general rules) or many (in orbifold constructions) of the terms present in such
superpotentials  are forbidden in the string framework.
Let us mention that, in spite of the severe constraints we are finding, many
of the most relevant terms needed in standard GUTs  do reappear here. This is
quite interesting, because this was by no means guaranteed from the beginning
and a completely different scenario could have emerged.
  For instance, in the model presented above  there is a $\underline {54}$
and $\bar {16}+ \underline {\bar {16}}$ and $\underline {10}$s,
needed for GUT  symmetry breaking down to
the standard model. There are couplings
\break {$\underline {10}$ $ \underline {10}$ $ \underline {16}$} needed for
fermion
masses, $\underline {16}_i$$ \underline {\bar {16}}$$ X_i$ ($i=1,\dots
N_{gen}$) with $X$
a singlet, which could be used for neutrino masses etc.
The fact that usually  four generation (or multiples of four) models are found,
is related to the kind of orbifolds we have been considering so far \break
($Z_2\times$ $Z_2,Z_4$ $Z_2\times Z_4 \dots$). Three generations would be
easier to be found in
orbifolds admitting order three Wilson lines ($Z_6 \dots$)\cite{afiu2}.

The other fundamental unanswered question in SusyGUTs  refers to the doublet
triplet splitting problem. Interestingly enough, most of the models found
posses terms of the form
\begin{equation}
W_X\ =\ \lambda
	H\Phi  {\bar H}\ +\ XH{\bar H}   \ .
\label{supx}
\end{equation}
where $\Phi=\underline {54}$  ( $\underline {24}$) $H=\underline {10}$ (
\underline 5 ) and $X$ a singlet of $SO(10)$ ($SU(5)$). For instance, in the
the model $SO(10)$ model presented above, it is possible to see
that there exist flat directions which break
the symmetry down to
$SU(4)\times SU(2)\times SU(2)$ with some of the doublets remaining light
whereas the colour triplets remain heavy.
It is well known that this mechanism is spoiled by quantum corrections in
field theory \cite{slids}. Studying its stability in string theory would
involve non perturbative physics determining preferred directions in moduli
space. Because of our present ignorance, we can not rule out this mechanism
in the string context. Recall that there is no place for a
missing partner mechanism
(at least for $k\le 5$) and that fine tuning is not even possible here since
there are no mass terms.


\begin{thebibliography}{99}
\bibitem{afiu}
G. Aldazabal, A. Font, L.E. Ib\'a\~nez and A.M. Uranga,
 Madrid preprint
FTUAM-94-28 (1994); hep-th/9410206. See also L. Ib\'a\~nez, talk at {\it
Strings 95},
USC, March 1995; hep-th/9505098.
%
\bibitem{afiu2}
G. Aldazabal, A. Font, L.E. Ib\'a\~nez and A.M. Uranga,
to appear.
%
\bibitem{amal}
 J. Ellis, S. Kelley and D.V. Nanopoulos,  Phys.Lett. B249 (1990)
441;   B260 (1991) 131; P. Langacker and M. Luo,  Phys.Rev.  D44
(1991) 817; U. Amaldi, W. de Boer and H. F\"urstenu,  Phys.Lett.  B260
(1991) 447. Several talks in this Susy $'95$ meeting.
%
\bibitem{iki}
L.E. Ib\'a\~nez,  Phys.Lett.  B318 (1993) 73.
%
\bibitem{fiq}
A. Font, L.E. Ib\'a\~nez and F. Quevedo,
Nucl. Phys. B345 (1990) 389.
%
\bibitem{lew}
D. Lewellen, Nucl. Phys. B337 (1990) 61.
%
\bibitem{otros}
S. Chaudhuri, S. Chung and J.D. Lykken, preprint
Fermilab-Pub-94/137-T, hep-ph/9405374;
S. Chaudhuri, S. Chung, G. Hockney and J.D. Lykken, preprint
hep-th/9409151.
G.B. Cleaver,
preprint OHSTPY-HEP-T-94-007; hep-th/9409096.
%
\bibitem{cchl}
S. Chaudhuri, S. Chung, G. Hockney and J.D. Lykken, preprint; hep-th/ 9501361.
\bibitem{ginsparg}
P. Ginsparg, Nucl. Phys. B295[FS21] (1988)153.
%
\bibitem{gepner}
D. Gepner, ``Lectures on $N=2$ String theory", Proceedings of Trieste
Spring School (1989), M.Green et al.(eds.), World Scientific; Singapur, 1990.
%
\bibitem{kz}
Y. Kazama and H. Suzuki, Nucl. Phys. B321 (1989)232.
%
\bibitem{fimqr}
A. Font, L.E. Ib\'a\~nez, M. Mondragon, F. Quevedo and G. G. Ross, Phys. Lett.
B227 (1989)34.
\bibitem{hall}
L. Hall and S. Raby, "A complete Supersymmetric $SO(10)$ Model", preprint
LBL-36357 (1995).
%
\bibitem{slids}
See e.g.  V. Kaplunovsky, Nucl.Phys.B233 (1984) 336.
%
\end{thebibliography}
\end{document}